\documentstyle[amssymb,aps]{revtex}

\begin{document}
\title{Deterministic Secure Direct Communication Using Ping-pong protocol without
public channel}
\author{Qing-yu Cai}
\address{Laboratory of Magentic Resonance and Atom and Molecular Physics, Wuhan\\
Institute of Mathematics, The Chinese Academy of Science, Wuhan,\\
430071,People's Republic of China}
\maketitle

\begin{abstract}
Based on an EPR pair of qubits and allowing asymptotically secure key
distribution, a secure communication protocol is presented. Bob sends either
of the EPR pair qubits to Alice. Alice receives the travel qubit. Then she
can encode classical information by local unitary operations on this travel
qubit. Alice send the qubit back to Bob. Bob can get Alice's information by
measurement on the two photons in Bell operator basis. If Eve in line, she
has no access to Bob's home qubit. All her operations are restricted to the
travel qubit. In order to find out which opeartion Alice performs, Eve's
operation must include measurements. The EPR pair qubits are destroyed.
Bob's measurement on the two photons in Bell operator basis can help him to
judge whether Eve exist in line or not. In this protocal, a public channel
is not necessary.

\end{abstract}

The idea of quantum cryptography was first proposed in the 1970s by Stephen
Wiesner [1] and by Charles H. Bennett of IBM and Gilles Brassard of The
University of Montr$e$al [5]. The motive that we build quantum-mechanical
communications channels is not only to transmit information securely without
being eavesdropped on but also to transmit information more efficiently. We
consider that quantum channel is secure because that quantum physics
establishes a set of negative rules stating things that cannot be done [2]:
(1) One cannot take a measurement without perturbing the system. (2) One
cannot determine simultaneously the position and the momentum of a particle
with arbitrarily high accuracy. (3) One cannot simultaneously measure the
polarization of a photon in the vertical-horizontal basis and simultaneously
in the diagonal basis. (4) One cannot draw pictures of individual quantum
processes. (5) One cannot duplicate an unknown quantum state. These
characters of quantum physics let us have ability to exchange information
securely.

As is well known, operations on one particle of an Einstein-Podolsky-Rosen
(EPR) pair cannot influence the marginal statistics of measurements on the
other particle. Let's suppose that Bob have two photons which are maximally
entangled in their polarization degree of freedom 
\begin{eqnarray}
|\psi ^{\pm } &>&=\frac{1}{\sqrt{2}}(|01>\pm |10>)  \nonumber \\
|\phi ^{\pm } &>&=\frac{1}{\sqrt{2}}(|00>\pm |11>)
\end{eqnarray}
These states are maximally entangled states in the two particle Hilbert
space $H=H_{A}\otimes H_{B}$. Each one of them is not polarized because the
reduced density matrices of one photon, say A, 
\begin{equation}
\rho _{A}^{\pm }=Tr_{B}\{|\psi ^{\pm }><\psi ^{\pm }|\}=Tr_{B}\{|\phi ^{\pm
}><\phi ^{\pm }|\}
\end{equation}
is complete mixture. Since $\rho _{A}^{\pm }=\frac{1}{2}$I$_{A}$, no one can
distinguish $|\psi ^{\pm }>$ (or $|\phi ^{\pm }>$) only by using the
information of photon A. Anyone who has these two photons can perform a
measurement on both photons in Bell operator basis which can help him to
distinguish the states $\{|\psi ^{+}>,|\psi ^{-}>,|\phi ^{+}>,|\phi ^{-}>\}$
from each other. Suppose Bob has two photons in Bell states $|\psi ^{\pm }>$
or $|\phi ^{\pm }>$, he gives one of them to Alice and keeps another. Then
Alice perform an operation which can be described by a unitary operator to
the photon she obtained from Bob 
\begin{eqnarray}
\stackrel{\wedge }{\sigma }_{Z}^{A}|\psi ^{\pm } &>&=|\psi ^{\mp }> 
\nonumber \\
\stackrel{\wedge }{\sigma }_{Z}^{A}|\phi ^{\pm } &>&=|\phi ^{\mp }>
\end{eqnarray}
where $\stackrel{\wedge }{\sigma }_{Z}^{A}\equiv (\stackrel{\wedge }{\sigma }%
_{Z}\otimes $I$)=(|0><0|-|1><1|)\otimes $I. The local operation of Alice can
be encoded in states $|\psi ^{\pm }>$. So it becomes nonlocal information.
We will give a secure communication protocal based on this. In order to
realize the communication, first, Bob prepares an EPR pair photons (suppose
in state $|\psi ^{+}>$). Then he sends one of them to Alice and keeps
another. Alice receives the photon and perform an operation in order to
encode her information. She can perform the operation $\stackrel{\wedge }{%
\sigma }_{Z}$ to encode her information ''1''. Else, she performs the
operation I (do nothing) to encode her information ''0'' . Alice sends the
encoded photon to Bob. Bob performs a measurement on both photons in Bell
operator basis to judge Alice's operation. If the EPR pair in state $|\psi
^{+}>$, then he know that Alice perform an operation I, else, he know that
Alice perform the operation $\stackrel{\wedge }{\sigma }_{Z}$. This is a $%
ping-pong$ $protocol$ [3]. Through this process, Bob can get one bit
information from Alice. When Bob's measurement with the result $|\phi ^{\pm
}>$, he knows that there is Eve in the line and stops the communication. We
will prove that if there is an Eve in the communication line, Bob can find
it certain without any public channel.

The algorithm for this protocol can like this: (1) Bob prepares two qubits
in Bell states $|\psi ^{+}>$ or $|\phi ^{+}>$ randomly. (2) Bob sends one of
the ERP pair to Alice and stores the other one. (3) Alice receives the $one$ 
$qubit$, to perform the her encoding operation. (4) Alice sends the qubit
back to Bob. (5) Bob receives the back qubit and performs a basis
measurement on both photons in Bell operator basis \{$|\phi ^{+}>,|\phi
^{-}>,|\psi ^{+}>,|\psi ^{-}>$\}. (6) Suppose Bob used $|\psi ^{+}>$ this
time, after his measurement, if he finds the two photons in state $|\psi
^{+}>$, then he know that Alice has encoded '0' in this process. If Bob find
two photons in state $|\psi ^{-}>$, then he knows that Alice has encoded '1'
in this precess. Then he will prepare next EPR pair and repeat this process.
Else, if Bob finds the two photons are not in states $|\phi ^{\pm }>$, then
he know that there is Eve in this communication line. The communication
stops. (If Bob use $|\phi ^{+}>$ this time, after his measurement, if he
finds the two photon in state $|\phi ^{+}>,$ he knows that Alice has encoded
'0' in this process. If he finds the two photons in state $|\phi ^{-}>$, Bob
knows Alice has encode '1' in this process. If Bob finds the two photons in
states $|\psi ^{\pm }>$, he knows Eve in line. The communication stops.) We
will prove that this process is secure.

$Security$ $proof$. Eve can use all technique quantum mechanics laws allows.
The aim of Eve is to find out which operation Alice performs. But Eve has no
chance to access Bob's home qubit. All her operations only can operator on
the travel qubit. Bob select EPR pair in states $|\phi ^{+}>$ or $|\psi ^{+}>
$ randomly with a probability $\frac{1}{2}$ every time. He send one of the
qubit to Alice. The state of the travel qubit is complete mixed to Eve 
\begin{equation}
\rho _{A}=Tr_{B}\{|\psi ^{+}><\psi ^{+}|\}=Tr_{B}\{|\phi ^{+}><\phi ^{+}|\}=%
\frac{1}{2}\text{I}_{A}.
\end{equation}
To Eve, it seems that Bob sends a qubit in state $|0>$ or $|1>$ with
probability $\frac{1}{2}$ every time. If Eve want to distinguish what
operations Alice operaters, she needs to attack the travel qubit before
Alice's operation. After Alice's operation, Eve needs to measure the bcak
qubit.(This process is discribed in figure 1 and 2.) Eve can use all
technique quantum mechanics allows. The most general quantum operation is a
completely positive map 
\begin{equation}
\varepsilon :S(H_{A})\rightarrow S(H_{A})
\end{equation}
one the state space $S(H_{A})$. Using the Stinespring dilation theorem [4],
we know that any completely positive map can be realized by a unitary
operation on a larger Hilbert space. We can use an ancilla space $H_{E}$ and
an ancilla state $|\chi >\in H_{E}$ and a unitary operation $\stackrel{%
\wedge }{E}$ on $H_{A}\otimes H_{E}$ to realize this completely positive map 
\begin{equation}
\varepsilon (\rho _{A})=Tr_{E}\{\stackrel{\wedge }{E}(\rho _{A}\otimes |\chi
><\chi |)\stackrel{\wedge }{E^{+}}\}
\end{equation}
where dim$H_{E}\leq $(dim$H_{A}$)$^{2}$. If Eve want to gain information
about Alice's operation, she should first perform the unitary operation $%
\stackrel{\wedge }{E}$ on the composed system, then let Alice perform her
coding operation on the travel qubit. After Alice's operation, Eve will
finally perform a measurement to judge which operation Alice performs. The
state of the travel qubit is a maximal mixture state 
\begin{equation}
\rho _{A}=\frac{1}{2}|0><0|+\frac{1}{2}|1><1|
\end{equation}
to Eve, which can be as a Bob sends the travel qubit in either of the states 
$|0>$ or $|1>$ with equal probability $p=\frac{1}{2}$.

In our proof, first we suppose Bob selects the EPR pair in state $|\psi ^{+}>
$. Second we suppose that Bob sends the travel qubit in state $|0>$ to
Alice. Eve adds an ancilla in the state $|\chi >$ and performs the unitary
operation $\stackrel{\wedge }{E}$ on both systems, which will results in 
\begin{equation}
|\psi _{AE}^{0}>=\stackrel{\wedge }{E}|0,\chi >=\alpha |0,\chi _{00}>+\beta
|1,\chi _{01}>
\end{equation}
where $|\chi _{0}>$, $|\chi _{1}>$ are pure ancilla states uniquely
determined by $\stackrel{\wedge }{E}$, and $|\alpha |^{2}+|\beta |^{2}=1$.
We change this equation into another form 
\begin{equation}
d=|\beta |^{2}=1-|\alpha |^{2}
\end{equation}
It has been proved in [3] that if Eve want to get $full$ information of
Alice's opeartion from the travel qubit, it must result in that $d=\frac{1}{2%
}$. So equation (8) can be write as 
\begin{equation}
|\psi _{AE}^{0}>=\frac{1}{\sqrt{2}}|0,\chi _{00}>+\frac{1}{\sqrt{2}}|1,\chi
_{01}>
\end{equation}
After Eve's attack opeartions, the travel qubit will be sent to Alice.
Suppose Alice perform an operation 
\begin{equation}
\stackrel{\wedge }{\sigma }_{Z}^{A}|\psi _{AE}^{0}>=(|0><0|-|1><1|)(\frac{1}{%
\sqrt{2}}|0,\chi _{00}>+\frac{1}{\sqrt{2}}|1,\chi _{01}>)=\frac{1}{\sqrt{2}}%
|0,\chi _{00}>-\frac{1}{\sqrt{2}}|1,\chi _{01}>.
\end{equation}
to encode the information '1'. After Alice's encoding operation, the joint
system that includes Bob's home qubit, the travel qubit and Eve's ancilla
space is in state 
\begin{equation}
|\psi _{BAE}^{0}>=\frac{1}{\sqrt{2}}|0_{A},\chi _{00},1_{B}>-\frac{1}{\sqrt{2%
}}|1_{A},\chi _{01},1_{B}>=(\frac{1}{\sqrt{2}}|0,\chi _{00}>-\frac{1}{\sqrt{2%
}}|1,\chi _{01}>)|1_{B}>  \eqnum{*1}
\end{equation}
And the system including the travel qubit and Eve's ancilla space is in the
state 
\begin{equation}
|\psi _{AE}^{0}>=\frac{1}{\sqrt{2}}|0,\chi _{00}>-\frac{1}{\sqrt{2}}|1,\chi
_{01}>  \eqnum{**1}
\end{equation}
After Alice's operation, Eve has to measure the travel photon. No
measurement implies that Eve cannot get any information about the travel
photon. In this process, if Alice' operation is I, then (*1) shoul be
written as 
\begin{equation}
|\psi _{BAE}^{0}>=(\frac{1}{\sqrt{2}}|0,\chi _{00}>+\frac{1}{\sqrt{2}}%
|1,\chi _{01}>)|1_{B}>  \eqnum{*2}
\end{equation}
and (**1) should be written as 
\begin{equation}
|\psi _{AE}^{0}>=\frac{1}{\sqrt{2}}|0,\chi _{00}>+\frac{1}{\sqrt{2}}|1,\chi
_{01}>  \eqnum{**2}
\end{equation}
At the beginning of our proof, we have supposed that Bob sends the travel
qubit in state $|0>$. If Bob sends the travel qubit in states $|1>$, (*1)
should be written as 
\begin{equation}
|\psi _{BAE}^{1}>=(\frac{1}{\sqrt{2}}|0,\chi _{00}>-\frac{1}{\sqrt{2}}%
|1,\chi _{01}>)|0_{B}>  \eqnum{*3}
\end{equation}
and (**1) should be written as 
\begin{equation}
|\psi _{AE}^{1}>=\frac{1}{\sqrt{2}}|0,\chi _{10}>-\frac{1}{\sqrt{2}}|1,\chi
_{11}>  \eqnum{**3}
\end{equation}
when Alice's operation is $\stackrel{\wedge }{\sigma }_{Z}$. And (*1) should
be written as 
\begin{equation}
|\psi _{BAE}^{1}>=(\frac{1}{\sqrt{2}}|0,\chi _{10}>-\frac{1}{\sqrt{2}}%
|1,\chi _{11}>)|0_{B}>  \eqnum{*4}
\end{equation}
(**1) should be written as 
\begin{equation}
|\psi _{AE}^{1}>=\frac{1}{\sqrt{2}}|0,\chi _{10}>-\frac{1}{\sqrt{2}}|1,\chi
_{11}>  \eqnum{**4}
\end{equation}
when Alice's operation is I. Eve can obtain information with certain by
measurement on the travel qubit in basis \{$\frac{1}{\sqrt{2}}(|0,\chi
_{00}>+|0,\chi _{01}>),$ $\frac{1}{\sqrt{2}}(|0,\chi _{00}>-|0,\chi _{01}>),$
$\frac{1}{\sqrt{2}}(|1,\chi _{10}>+|1,\chi _{11}>),$ $\frac{1}{\sqrt{2}}%
(|1,\chi _{10}>-|1,\chi _{11}>)$\}. After Eve's measurement, the travel
qubit will be sent to Bob. Bob does not know whether Eve exist or not, so
there is 
\begin{equation}
\rho _{AB}=Tr_{E}(|\psi _{ABE}><\psi _{ABE}|).
\end{equation}
He measure the two photons in Bell basis, then 
\[
Tr(|\psi ^{+}><\psi ^{+}|\rho _{AB})=\frac{1}{4}
\]
\[
Tr(|\psi ^{-}><\psi ^{-}|\rho _{AB})=\frac{1}{4}
\]
\[
Tr(|\phi ^{+}><\phi ^{+}|\rho _{AB})=\frac{1}{4}
\]
\begin{equation}
Tr(|\phi ^{-}><\phi ^{-}|\rho _{AB})=\frac{1}{4}
\end{equation}
If Eve does not exist, which means $d=0$, we can see when Alice performs
operation I, there is 
\[
Tr(|\psi ^{+}><\psi ^{+}|\rho _{AB})=1
\]
\[
Tr(|\psi ^{-}><\psi ^{-}|\rho _{AB})=0
\]
\[
Tr(|\phi ^{+}><\phi ^{+}|\rho _{AB})=0
\]
\begin{equation}
Tr(|\phi ^{-}><\phi ^{-}|\rho _{AB})=0
\end{equation}
When Alice operator $\stackrel{\wedge }{\sigma }_{Z}^{A}$, Bob's measurement
result should be 
\[
Tr(|\psi ^{+}><\psi ^{+}|\rho _{AB})=0
\]
\[
Tr(|\psi ^{-}><\psi ^{-}|\rho _{AB})=1
\]
\[
Tr(|\phi ^{+}><\phi ^{+}|\rho _{AB})=0
\]
\begin{equation}
Tr(|\phi ^{-}><\phi ^{-}|\rho _{AB})=0
\end{equation}
We can see that we have supposed that Bob used the EPR pair in state $|\psi
^{+}>$ in this process. It is apparent that when Eve exist, Bob will find
his measurement result will be in the states $|\phi ^{\pm }>$ with
probability $\frac{1}{2}$. In another word, when Bob uses an EPR pair in
states $|\psi ^{+}>$ in the communication but he find his measurement result
is in states $|\phi ^{\pm }>$, he knows Eve is in line. The communication
stops. For the same reason, when Bob select an EPR pair in state $|\phi ^{+}>
$ to communicate with Alice but he finds his measurement result is in the
state $|\psi ^{\pm }>$, he knows that Eve is in line. The communication
stops.

If Bob only uses EPR pair in state $|\psi ^{+}>$ in the communication, Eve
can perform a measurement in her attack operations to determine whether Bob
sends the travel qubit in states $|0>$ or $|1>$. After her measurement
gained Alice's operation information, she can prepare the travel qubit in
the states $|0>$ or $|1>$ as it is at the beginning. When Bob selects EPR
pair in states $|\psi ^{+}>$ or $|\phi ^{+}>$ randomly with a probability p=$%
\frac{1}{2}$, Eve can not determine which states Bob select by her
measurement on the travel qubit. We have proved that this communication
protocol is secure.

We can see that when Eve wants to gain full information in each attack, the
ping-pong protocol provides a detection probability $P=\frac{1}{2}$. If Eve
exist, after 1000 bits have been transmitted, the probability that Eve was
not detected becomes $D\thickapprox 9.33e-302$, which means Eve has already
been detected. In fig. 3, we have plotted the eavsdropping sucess
probability with $d=\frac{1}{2}$.

In our method, every travel photon can get one bit information back from
Alice without Eve in line. Bob's measurement on the two photons in Bell
operator basis not only can help him to get information from Alice, but also
can can help him to determine whether Eve is in line or not. Obviously,
there is no public channel in our method. In BB84's protocol [5], the states
some photons have to be publicized to verdict the communication is secure.
In the ping-pong protocol [3], some EPR pair photons have to be publicized
to judge whether Eve exist in line or not through the public channel. In a
secure communication protocal, first, an additionl public channel is not
economical. On the other hand, a public channel means all information
through the public channel is open to Eve. The more information Eve gains,
the more difficult the secure communication becomes.

\subsubsection{References:}

[1]. S. Wiener, then at Columbia University, was first to propose ideas
closely related to QC in the 1970s. Since it is difficult to find his
revolutionary paper, we reproduce his abstract here: The uncertainty
principle imposes restrictions on the capacity of certain types of
communication channels. This paper will show that in compensation for this
'quantum noise,' quantum mechanics allows us novel forms of coding without
analogue in communication channels adequately described by classical physics.

[2]. Nicolas Gisin, Gr$\stackrel{^{\prime }}{e}$goire Ribordy, Wolfgang
Tittel, and Hugo Zbinden, Rev. Mod. Phys. 74, 145 (2002).

[3]. Kim Bostr$\stackrel{..}{o}$m and Timo Felbinger, Phys. Rev. Lett. 89,
187902 (2002).

[4]. W. F. Stinespring, Proc. Am. Math. Soc. 6, 211 (1955).

[5]. C. H. Bennett and G. Brassard, 1984, in $proceedings$ $of$ $the$ $IEEE$ 
$International$ $Conference$ $on$ $Computers$, $Systems$ $and$ $%
\mathop{\rm Si}%
gnal$ $\Pr oces\sin g$, Bangalor, India, (IEEE, New York), pp. 175-179.

\end{document}